\documentclass[10pt,oneside]{paper}

\usepackage[a4paper, total={7in, 9in}]{geometry}

\usepackage[ruled,linesnumbered]{algorithm2e}
\makeatletter

\usepackage{amsmath,amssymb}
\usepackage{cite}

\usepackage{nameref,hyperref}
\date{}

\usepackage{lastpage,fancyhdr,graphicx}

\title{NGS Based Haplotype Assembly Using Matrix Completion} 
\author{Sina Majidian , Mohammad Hossein Kahaei}
\begin{document}
\maketitle

\section*{Abstract}

We apply matrix completion methods for haplotype assembly from NGS reads to develop the new HapSVT, HapNuc, and HapOPT algorithms. This is performed by applying a mathematical model to convert the reads to an incomplete matrix and estimating unknown components. This process is followed by quantizing and decoding the completed matrix in order to estimate haplotypes. These algorithms are compared to the state-of-the-art algorithms using simulated data as well as the real fosmid data. It is shown that the SNP missing rate and the haplotype block length of the proposed HapOPT are better than those of HapCUT2 with comparable accuracy in terms of reconstruction rate and switch error rate. A program implementing the proposed algorithms in MATLAB is freely available at https://github.com/smajidian/HapMC.


\section*{Introduction}
The Single Nucleotide Polymorphism (SNP) is a kind of genetic variation with a  frequency greater than 1\% in population. In diploid organisms, genomes are organized into pairs of chromosomes, a paternal and a maternal copy. The sequence of SNPs on each copy of a pair of chromosomes is called a haplotype. A genotype is the conflation of two haplotypes on the homologous chromosomes. An SNP is called homozygous, if a pair of alleles at this locus is made up of two identical nucleotides, and  is heterozygous, otherwise.

From the evolutionary point of view, the SNP happens as a consequence of mutation. However, since the  mutation rate is low, several mutations of a locus rarely occur. Thus, it is usual to assume that the majority of SNPs are bi-allelic, meaning that each SNP can be chosen from just two of the four possible nucleotides, $ i.e.$, A, T, C, and G \cite{del}.  Accordingly, in this work we similarly use this assumption.  The haplotype is widely used in the  Genome Wide Association Studies (GWAS), clinical genetics, linkage analysis, drug-design, and personalized medicine \cite{sny}.

To extract a haplotype, one may use the following three approaches where the last two approaches are mathematical:\\

\noindent 1) Applying high-cost experimental and expensive methods for every single individual which is of course not desirable \cite{sny}. \\

\noindent 2) Haplotype phasing wherein the haplotypes are inferred from the genotypes of multiple individuals. As such, a method based on the maximum parsimony assumption \cite{wan} and statistical methods like SHAPEIT, developed based on the Hidden Markov Model \cite{del, oc} may be mentioned. Note that using this approach, the haplotype of an individual can not be found separately and also is challenged by the low-frequency and also \emph{de novo} variants  \cite{sny}. \\

\noindent 3) Estimating haplotypes from  Next Generation Sequencing (NGS) reads $i.e.$  nucleotide sequence of fragments. Using this approach,  known as the haplotype assembly, haplotyping of a single individual becomes feasible. In this regard, HapCUT2 \cite{hapcut}, HapTree \cite{haptree}, and HapSAT \cite{hapsat} are three famous methods developed based on probabilistic models. These methods are sensitive to the selected model and thus fragile to the model error.

A recent method for haplotype assembly is AltHap \cite{hash} which has shown accurate results compared to H-PoP \cite{xie16}, SCGD \cite{ccai}, and HapTree \cite{haptree}. The H-PoP is a heuristic algorithm originated from the Balanced Optimal Partition (BOP) optimization model which benefits from the Minimum Error Correction (MEC) as well as the maximum fragments cut approaches \cite{xie12}.
The SDhaP \cite{das} is also another heuristic method based on correlation clustering and non-convex optimization which does not guarantee reaching the global optimum.

The innovation of this article is threefold. First, the haplotype assembly is mathematically formulated based on matrix completion methods.
Secondly,  three new algorithms called the Haplotype assembly based on Singular Value Thresholding (HapSVT), Haplotype assembly based on Nuclear norm minimization (HapNuc), and Haplotype assembly based on OPTSPACE (HapOPT) are proposed. Next, in the section of Results, these algorithms are compared to some benchmark methods in terms of the reconstruction rate and the switch error rate.

\section*{Model of Haplotypes}
To exploit the NGS reads as the raw data, a computational modeling is needed. For this purpose, similar to \cite{ccai}, we first  convert the sequence of nucleotides which can be either reads or haplotypes into a sequence of numbers.
The SNP nucleotides are converted to $1$ and $-1$ for the wild and rare alleles, respectively. As an example, Table 1 depicts the alleles of the $\beta_2$AR gene  \cite{wan} for which the maternal and paternal haplotypes of an individual are shown by $\boldsymbol h_m$ and $\boldsymbol h_p$, respectively. The corresponding codewords based on the above modeling are presented in the last column.
\begin{table}[!ht]
\centering
\caption{ {\bf Haplotypes of $\beta_2$AR genes and their corresponding codewords.}}
{\begin{tabular}{|l|cccccccccc|c|}\hline
        & \multicolumn{10}{|c|}{Nucleotides}& Codewords \\\hline
Alleles & G/A& C/A& G/A& C/G& T/C& T/C& T/C& G/A& C/G& G/A& \{1/-1,1/-1,...\}\\ \hline
$\boldsymbol h_m$& A& C& G& G& C& C& C& G& G& G&\{-1,1, 1,-1,-1,-1,-1, 1,-1,1\}\\
$\boldsymbol h_p$& G& C& A& C& T& T& T& A& C& G&\{ 1,1,-1, 1, 1, 1, 1,-1, 1,1\} \\ \hline  \end{tabular}}
\label{table1}
\end{table}

Next, assuming that each read has been aligned to the reference genome, the non-SNP sites of each read are omitted.
Then, the reads are coded using the procedure described in Table 1, and are completed by adding zeros for the length of $l$ as shown for 10 aligned  reads in Table 2. As seen in this example,  for the $1$st row, we get  \{-1 1 1 0 0 0 0 0 0 0\} with 3 sites of  $\pm 1$ and  7 sites of zeros.

\begin{table}[!h]
\centering
\caption{ {\bf Example of aligned reads for $\beta_2$AR genes and the considered codewords.}}
{\begin{tabular}{|c|cccccccccc|cccccccccc|}\hline
Reads & \multicolumn{10}{|c|}{Nucleotides}&\multicolumn{10}{|c|}{Codewords}\\\hline
1& A&C& G&&&&&&&& -1& 1&  1&  0&  0&  0&  0&  0&  0& 0\\
2&&& G& G& C& C&&&&& 0& 0&  1& -1& -1& -1&  0&  0&  0& 0\\
3&&& G& G&&&&&G&G& 0& 0&  1& -1&  0&  0&  0&  0& -1& 1\\
4& G&C& A& C& T& T&&&&& 1& 1& -1&  1&  1&  1&  0&  0&  0& 0\\
5&&& A& C&&& T& A& C&G& 0& 0& -1&  0& 0&  1&  1& -1&  1& 1\\
6& G&C&&& T& T&&&&& 1& 1&  0&  0&  1&  1&  0&  0&  0& 0\\
7&&C&&& C&&&&&G& -1& 1&  0&  0& -1&  0&  0&  0&  0& 1\\
8& A&C&&& C& C& C&&&& -1& 1&  0&  0& -1& -1& -1&  0&  0& 0\\
9& G&&& C&&& T& A& C&& 1& 0&  0&  1&  0&  0&  1& -1&  1& 0\\
10&&& A& C&&&&&C&G& 0& 0& -1&  1&  0&  0&  0&  0&  1& 1\\\hline
\end{tabular}}
\label{table1}
\end{table}

Without loss of generality, by representing the codewords of Table 2 by the vectors $\boldsymbol r_i, i=1,...,N$, we form the read matrix $\boldsymbol R$, where $N$ is the number of reads. In fact, $\boldsymbol R$ is an incomplete matrix with the rank of 2 which consists of the maternal and paternal haplotypes in its rows. At this stage,  we may utilize matrix completion methods to complete this low rank matrix. To do so, by estimating the zero entries of $\boldsymbol R$, we obtain the completed matrix $\boldsymbol H$ which has the same dimension as  $\boldsymbol R$, $i.e.$, $N \times l$ where $l$ is the haplotype length. According to Table 2, these matrices  are given by (1) and (2).
\begin{equation}
\boldsymbol R=
\begin{bmatrix}
-1& 1&  1&  0&  0&  0&  0&  0&  0& 0 \\
0& 0&  1& -1& -1& -1&  0&  0&  0& 0 \\
0& 0&  1& -1&  0&  0&  0&  0& -1& 1 \\
1& 1& -1&  1&  1&  1&  0&  0&  0& 0 \\
0& 0& -1&  0&  0&  1&  1& -1&  1& 1 \\
1& 1&  0&  0&  1&  1&  0&  0&  0& 0 \\
-1& 1&  0&  0& -1&  0&  0&  0&  0& 1 \\
-1& 1&  0&  0& -1& -1& -1&  0&  0& 0 \\
1& 0&  0&  1&  0&  0&  1& -1&  1& 0 \\
0& 0& -1&  1&  0&  0&  0&  0&  1& 1 \\
\end{bmatrix}
\end{equation}
\begin{equation}
\boldsymbol H=\begin{bmatrix}
-1& 1& 1& -1& -1& -1& -1& 1& -1& 1\\
-1& 1& 1& -1& -1& -1& -1& 1& -1& 1\\
-1& 1& 1& -1& -1& -1& -1& 1& -1& 1\\
 1& 1& -1& 1& 1& 1& 1& -1& 1& 1\\
 1& 1& -1& 1& 1& 1& 1& -1& 1& 1\\
 1& 1& -1& 1& 1& 1& 1& -1& 1& 1\\
-1& 1& 1& -1& -1& -1& -1& 1& -1& 1\\
-1& 1& 1& -1& -1& -1& -1& 1& -1& 1\\
 1& 1& -1& 1& 1& 1& 1& -1& 1& 1\\
 1& 1& -1& 1& 1& 1& 1& -1& 1& 1\\
\end{bmatrix}
\end{equation}

From $\boldsymbol H$, one can observe that only two of its rows are different and thus the desired haplotypes are given by
\begin{equation}
\boldsymbol h_m=\begin{bmatrix}
-1& 1& 1& -1& -1& -1& -1& 1& -1& 1\\
\end{bmatrix},
\end{equation}
\begin{equation}
\boldsymbol h_p=\begin{bmatrix}
1& 1& -1& 1& 1& 1& 1& -1& 1& 1\\
\end{bmatrix}.
\end{equation}
These vectors can then be decoded to the sequence of nucleotides using the first row of Table 1. To the best of our knowledge, no algorithm has been reported to distinguish between the maternal and paternal haplotypes and therefore  $\boldsymbol h_p$ and $\boldsymbol h_m$ may be interchanged with each other.

It should be noted that the above example is an error-free case to clarify the procedure of data modeling which can be trivially solved. For the erroneous case; which is the subject of our work, $\boldsymbol R$ is an incomplete version of $\boldsymbol H + \boldsymbol N$ where $\boldsymbol N$ shows the noise matrix \cite{hash}.

\section*{Proposed Methods}
We present three new algorithms for haplotype assembly whose general block diagram is illustrated in Fig 1. The goal is to estimate $\boldsymbol h_p$ and $\boldsymbol h_m$  from the noisy reads. The first two blocks have been explained before.
\begin{figure}[!h]
  \centering
    \includegraphics{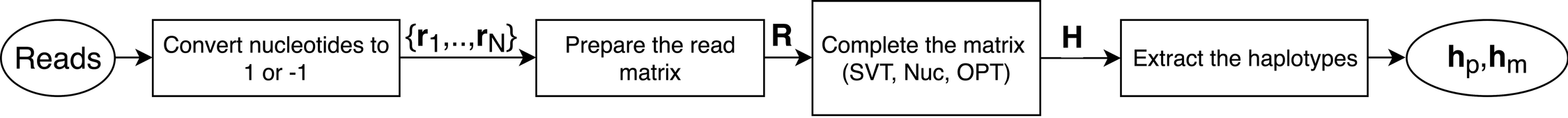} 
\caption{{\bf Block diagram of the proposed algorithms. }}
\label{fig1}
\end{figure}
In the third block, we receive an incomplete matrix $\boldsymbol R$ with a few known entries where the set of indices of known entries is given by $\Omega$ \cite{ccai}. Then, we intend to estimate the unknown entries based on rank assumption. Mathematically, this is modeled by the following optimization problem:
\begin{equation}
\min_{\substack{\boldsymbol H}}  \sum_{(i,j)\in \Omega} (H_{ij}-R_{ij})^2 \text{ subject to }  \text{rank} (\boldsymbol H) =2.
\end{equation}
It is worth mentioning that here we have not only considered the case of all-heterozygous variants, but also included the case of both heterozygous and homozygous variants. This can be realized as a point of this work in comparison to some other methods that are restricted to heterozygous variants. In the all-heterozygous case, the two haplotypes will be the negative of each other, $i.e.$, $\boldsymbol{h}_p=-\boldsymbol{h}_m$ and thus the rank of $\boldsymbol{H}$ will be one (See (5)).

To solve (5), the nuclear norm minimization, Singular Value Thresholding (SVT), and OPTSPACE methods have already been reported \cite{dav}, based on which we introduce three new algorithms called the HapSVT, HapNuc, and HapOPT.

\subsection*{Haplotype assembly based on Singular Value Thresholding (HapSVT)} %

To explain the proposed HapSVT algorithm, we first introduce the SVT which is based on Singular Value Decomposition (SVD) \cite{jcai} defined for the read matrix $\boldsymbol R$ as
\begin{equation}\label{svd}
\boldsymbol R= \boldsymbol U \boldsymbol\Sigma \boldsymbol V^H ,  \quad \boldsymbol\Sigma=\text{diag}(\sigma_i) \quad i=1,...,r
\end{equation}
where $H$ denotes the hermitian operator, and $\boldsymbol U$ and $\boldsymbol V$ have orthonormal columns with the dimension of $N \times r$ and $l \times r$, respectively.
By applying the singular value shrinkage operator $D_\tau(\cdot)$ to $\boldsymbol R$, we obtain
\begin{equation}\label{shrink}
D_\tau (\boldsymbol R)= \boldsymbol U D_\tau(\boldsymbol \Sigma) \boldsymbol V^H, \end{equation}
where
\begin{equation}\label{Dtau}
D_\tau(\boldsymbol\Sigma)=\text{diag}\big(\text{max}\{\sigma_i-\tau,0\}\big) .
\end{equation}
It is worth noting that $D_\tau (\boldsymbol R)$ is the optimal value of the optimization problem
\begin{equation}
\min_{\substack{\boldsymbol Z}}   \frac{1}{2}  \|\boldsymbol{R}-\boldsymbol Z\|^2_F +\tau \|\boldsymbol Z\|_*,
\end{equation}
where $ \|\cdot\|_F$ is the Frobenius norm and  $ \|\cdot\|_*$ shows the nuclear norm as the  summation of singular values.

To perform the matrix completion part as shown in Fig 1, we  recursively use the SVT  in  two steps. In the first step, starting with the initial matrix $\boldsymbol Y^0 = \boldsymbol R$, the singular value shrinkage operator is used as

\begin{equation}
\boldsymbol X^k=D_\tau (\boldsymbol Y^{k-1}) .
\end{equation}

Then, in the second step, the difference between the projected matrix $\boldsymbol X^k$ and the initial matrix is compensated  for the known entries using
\begin{equation}\label{step2}
\boldsymbol Y^k=\boldsymbol Y^{k-1}+\delta \mathcal{P}_\Omega(\boldsymbol R-\boldsymbol X^k),
\end{equation}
for $k=1,2, \ldots, $ where  $\mathcal{P}_{\Omega}(\cdot)$ is an operator which keeps the entries of the matrix corresponding to $\Omega$ unchanged, and sets the other entries to zero.
The iterations continue until the condition $\| \mathcal{P}_{\Omega}(\boldsymbol X^k - \boldsymbol  R)\|_F<\epsilon \| \boldsymbol R\|_F$ is satisfied and the last $\boldsymbol X^k $ is reported as the completed matrix $\boldsymbol{H}$.

To extract $\boldsymbol h_p$ and $\boldsymbol h_m$, we compute the reduced row echelon form of $\boldsymbol{H}$ and by using the first two pivot positions, two independent rows of $\boldsymbol{H}$ are obtained. Then, in order to acquire the paternal and maternal haplotypes the entries are quantized to $1$ and $-1$.
The procedures of the HapSVT algorithm is depicted in Algorithm 1.

\begin{algorithm}[H]
\SetAlFnt{\small}
\SetAlgoLined
\SetNoFillComment
\SetKwInOut{Input}{input}\SetKwInOut{Output}{output}
\Input{$N$ aligned reads}
\Output{Haplotypes  $\boldsymbol h_m,\boldsymbol h_p$}
\tcc{Read Matrix Preparation}
Convert the sequences of nucleotides (reads) to the sequences of numbers. \\
Add zeros to each read to construct $\boldsymbol r_i$s with the length of $l$.\\
Construct the read matrix $\boldsymbol R$ ($N\times l$).\\
\tcc{Matrix Completion (SVT)}
Initialize $Y^{0}=\boldsymbol R$, $k=0$, $i=1$.\\
\While{$\| \mathcal{P}_{\Omega}(\boldsymbol X^k - \boldsymbol  R)\|_F<\epsilon \| \boldsymbol R\|_F$}{$ k=k+1$\\
  	$\boldsymbol X^{k}=D_\tau (\boldsymbol Y^{k-1})$\\
	$\boldsymbol Y^{k}=\boldsymbol Y^{k-1}+\delta \mathcal{P}_{\Omega}(\boldsymbol R-\boldsymbol X^k)$}
	$\boldsymbol H= \boldsymbol X^{k}$\\
\tcc{Reduced Row Echelon Form (RREF) Calculation}
$[\boldsymbol{H}_r, \boldsymbol{p}]=\textrm{RREF}(\boldsymbol H^T)$\\
\tcc{Haplotype Extraction}
$\boldsymbol{H}_q=2*(\boldsymbol{H}>0)-1 $\\
$\boldsymbol{h}_p=\boldsymbol{H}_q(\boldsymbol{p}(1),:) $\\
$\boldsymbol{h}_m=\boldsymbol{H}_q(\boldsymbol{p}(2),:) $\\
Convert the entries of $\boldsymbol{h}_m$ and $\boldsymbol{h}_p$ to the nucleotides.
\caption{Haplotype assembly using SVT (HapSVT).}
\end{algorithm}

\subsection*{Haplotype assembly based on Nuclear norm  minimization (HapNuc)}

A popular method for matrix completion is based on relaxing the non-convex rank function to a convex function. Since the number of nonzero singular values determines the rank of a matrix, an approximation of the rank function is defined by the summation of singular values, known as the nuclear norm \cite{can10}. In this way, the optimization problem is cast as
\begin{equation}
\min_{\substack{\boldsymbol H}} \|\boldsymbol H\|_*  \quad \text{ subject to }   \| \mathcal{P}_{\Omega}(\boldsymbol{H} - \boldsymbol{R})\|_F<\epsilon.
\end{equation}

This problem can be solved easily using the CVX, a MATLAB based package \cite{cvx}.
It has been shown that the nuclear norm minimization has strong mathematical guarantees to achieve the optimal solution \cite{can10, can09, rec}. To develop the new HapNuc algorithm, we substitute the SVT part of Algorithm 1 by nuclear norm minimization.

\subsection*{Haplotype assembly based on OPTSPACE (HapOPT)} %

Another method for  matrix completion is known as OPTSPACE \cite{kesh}  in which unlike the two previous methods, we assume that the rank of the desired matrix $\boldsymbol H$ is  known. The OPTSPACE consists of the following three steps: a) trimming, b) projection, and c) cleaning, as explained below.\\
\noindent a) In the trimming step, those columns of $\boldsymbol R$ with the degrees larger than $2|\Omega|/l$ are set to zero where $|\cdot|$ shows the cardinality of a set and $l$ is the haplotype length. The degree of a column (or a row) shows the number of its known entries. This step is also performed for the rows of $\boldsymbol R$ with the degrees larger than $2|\Omega|/N$ where $N$ is the number of reads.\\
\noindent b) The trimmed  $\boldsymbol R$ obtained from Step (a)  is projected to the space of rank $r$ matrices using
\begin{equation}
P(\boldsymbol R)=\frac{Nl}{|\Omega|}  \boldsymbol UP_r(\boldsymbol\Sigma) \boldsymbol V^H,
\end{equation}
 where $P_r(\boldsymbol \Sigma)=\textrm{diag}(\sigma_1,...\sigma_r)$ and  $\boldsymbol U$  and $\boldsymbol V$ are given by \eqref{svd}. \\
\noindent c) The cleaning step is performed by solving the following optimization problem,
\begin{equation}
\min_{\substack{\boldsymbol X\in\mathbb{R}^{N\times r}, \boldsymbol Y\in\mathbb{R}^{l\times r} }}  \min_{\substack{\boldsymbol S \in\mathbb{R}^{r\times r}}}   \sum_{(i,j) \in \Omega} \big(\boldsymbol R_{ij}- (\boldsymbol X \boldsymbol S \boldsymbol Y^H)_{ij}\big)^2,
\end{equation}
which contains two minimization parts. The  inner part results in a function in terms of  $\boldsymbol X$ and $ \boldsymbol Y$. To solve the outer minimization part, we use a gradient based recursive method whose initial matrices are  computed from Step (b), $i.e.,$ $\boldsymbol X_0=\boldsymbol U$ and $\boldsymbol Y_0=\boldsymbol V$. Then, this recursive method leads to the optimal solution   $\boldsymbol{H}=\boldsymbol{X}_{\text{opt}} \boldsymbol S_{\text{opt}} \boldsymbol Y_{\text{opt}}^H$.
To finalize the third new HapOPT algorithm, we should substitute the SVT part of Algorithm 1 by the above three steps.

\section*{Results}

Using extensive simulations, we compare the performance of the proposed HapSVT, HapNuc, and HapOPT algorithms with that of the three recent benchmark algorithms AltHap \cite{hash}, HapCUT2 \cite{hapcut}, and SDhaP \cite{das}. It has already been shown that these algorithms outperform some other algorithms like  RefHap \cite{refhap}, SCGD \cite{ccai}, HapTree \cite{haptree}, and H-PoP \cite{xie16}.
For comparison purposes, a well-known criterion is the reconstruction rate defined as \cite{ger}
\begin{equation}
\text{rr}=1-\frac{1}{l}\text{min}\Big\{ \mathcal{HD}\big(\hat{\boldsymbol h}_m,\boldsymbol h_m \big),\mathcal{HD}\big(\hat{\boldsymbol h}_p,\boldsymbol h_p\big)\Big\},
\end{equation}
where $\hat{\boldsymbol h}_p$ and $\hat{\boldsymbol h}_m$ are the reconstructed haplotypes which are compared to the known maternal and paternal haplotypes, $\boldsymbol h_m$ and $\boldsymbol h_p$. Moreover, $\mathcal{HD}(\cdot,\cdot)$ is the augmented hamming distance between two vectors which counts the number of non-identical sites using
\begin{equation}
 \mathcal{HD}(\boldsymbol a,\boldsymbol b)= \sum_{j=1}^{l}{ \mathcal{D} \big(\boldsymbol a(j),\boldsymbol b(j)\big)},
\end{equation}
where $\mathcal{D}(\cdot,\cdot)$ is defined as
\begin{equation}
 \mathcal{D}(a,b)=  \left\{\begin{array}{l}
0 \quad  a=b\\
1 \quad \text{otherwise}. \end{array}\right.
\end{equation}

To consider another criterion for performance evaluation, we make use of the SWitch Error Rate (SWER), defined as the number of switches divided by the haplotype length \cite{kul}. A switch happens when the parental origin of an allele with respect to that of the previous allele differs from one parent to another.  For example, by considering $\boldsymbol{h}_p= [1, 1, 1, 1]$ and $\boldsymbol{h}_m=[-1, -1, -1, -1]$  as the grand truth haplotypes and the estimated haplotypes as $\hat{\boldsymbol{h}}_p =[1, 1, -1, -1]$ and $\hat{\boldsymbol{h}}_m =[-1, -1, 1, 1]$, one switch has been occurred.

\subsection*{Simulated data}

First, we use the simulated data \cite{ger}  generated based on real human haplotypes in the HapMap project. This dataset; which contains different read matrices with various error rates and coverage values originated from different haplotype lengths, has vastly been used in previous studies \cite{deng, ccai, chen}. We choose the longest available haplotype from the dataset with the length of l = 700. The coverage value of the NGS paired-end reads varies from c = 3 to its greatest value c = 10. The average number of reads are N = 561, 936, and 1873 for coverage values of c = 3, 5, and 10, respectively. The number of SNPs covered in each read is a constant value equal to 7.4. Also, 10\% (and 20\%) of the entries of the read matrix are contaminated by noise with uniform distribution. The results are averaged over 100 independent trials of the experiment.

Table 3 shows the reconstruction rates for different coverage values and error rates. The corresponding SWERs are also depicted in Table 4. In this case, 
HapCUT2 is not examined, since it needs the Variant Call Format (VCF) file which is not available for this simulated dataset \cite{ger}. As seen in both Tables 3 and 4, the proposed HapOPT algorithm outperforms the others in terms of the reconstruction rate as well as the SWER.  It is worth reminding that the SDhaP solves a non-convex optimization problem using a heuristic technique with the gradient descent algorithm which does not guarantee reaching the global optimum. Furthermore, as a consequence of increasing the coverage value, a better performance is achieved by a lower SWER and a higher reconstruction rate.


\begin{table}[!ht]
\centering
\caption{ {\bf Reconstruction rates for different algorithms on simulated data \cite{ger}. The best values are in boldface.}}
{\begin{tabular}{|c|c|c|c|c|c|c|} \hline
coverage  & error rate (\%)  &  SDhaP & AltHap & HapOPT(Proposed) & HapSVT(Proposed) & HapNuc(Proposed)\\\hline
3  &10 &97.87 &99.04	 &\textbf{99.07}	&98.38	&98.32 \\\hline
5  &10 &99.19  &99.66	 &\textbf{99.72}	&97.21	&98.82 \\\hline
10 &10 &99.64  &\textbf{1}    &\textbf{1}		&99.53	&99.64\\\hline
3  &20 &96.66   &97.32	 &\textbf{97.38}	&97.00		&97.31\\\hline
5  &20 &97.36 &98.24	 &\textbf{98.43}	&97.47	&97.47\\\hline
10 &20 &97.02  &\textbf{99.45} 		&99.25	&98.66	&98.6\\\hline
\end{tabular}}
\label{table1}
\end{table}

\begin{table}[!ht]
\centering
\caption{ {\bf SWERs for different algorithms on simulated data \cite{ger}. The best values are in boldface. }}
{\begin{tabular}{|c|c|c|c|c|c|c|} \hline
coverage  & error rate (\%)  &  SDhaP & AltHap &HapOPT (Proposed) & HapSVT (Proposed) & HapNuc (Proposed)\\\hline
3  &10		&0.070	&0.038	&\textbf{0.027}	&0.111	&0.120\\\hline
5  &10		&0.019	&0.0058	&\textbf{0.004}	&0.207	&0.049\\\hline
10 &10 		&0.0018	&\textbf{0}		&\textbf{0}		&0.012	&0.003\\\hline
3  &20		&0.227 	&0.247	&\textbf{0.218}	&0.350	&0.345\\\hline
5  &20		&0.136	&0.123	&\textbf{0.101}	&0.243	&0.266\\\hline
10 &20		&0.065	&\textbf{0.0178}	&0.018	& 0.080	&0.121\\\hline
\end{tabular}}
\label{table1}
\end{table}

\subsection*{Real fosmid data}
We evaluate the proposed algorithms on the sequence data of the individual NA12878 fabricated based on a fosmid approach \cite{refhap}. The coverage of this data set is $c=3$ and the average read length is 40 kb, and hence, is a low-coverage and long-read dataset. For evaluation purposes, we consider the trio-phased haplotype from the GATK resource bundle, as the grand truth containing 1.3 million heterozygous variants in common with fosmid dataset \cite{kul, dep}. This dataset has already been used in several studies \cite{hapcut, kul, hash}.

In the simulated dataset used in the last section, each read overlaps at least one another read, while for the real data these overlaps do not necessarily occur. In this situation, our algorithm incorporates the overlaps for haplotype estimation, and as a result, the output of each algorithm is some disjoint parts of the whole haplotype, called haplotype blocks. To evaluate a common length for these blocks, we consider their mean and also the AN50 defined as the median of blocks lengths in base pairs weighted by a proportion of correctly estimated alleles \cite{haptree}. Also, we define the SNP Missing Rate (SMR) for each chromosome as the ratio of the number of missing SNPs in the estimates and the haplotype length \cite{mot}. The results on the real fosmid data are shown in Table 5. One can see that both HapOPT and AltHap algorithms achieve lower SNP missing rates in comparison to HapCUT2 and SDhaP. Moreover, HapOPT and AltHap have a better span in terms of AN50.

\begin{table}[!h]
\centering
\caption{ {\bf Mean and AN50 of haplotype blocks lengths for different algorithms on real fosmid data.}}
{\begin{tabular}{|c |c|c|c |c|c|c |c|c|c |c|c|c |} \hline
 &  \multicolumn{3}{c|}{SDhaP} & \multicolumn{3}{c|}{HapCUT2} &  \multicolumn{3}{c|}{AltHap}&  \multicolumn{3}{c|}{HapOPT (Proposed)} \\\hline
Chr. &SMR&  Mean & AN50 (kb) & SMR& Mean& AN50 (kb) & SMR& Mean& AN50 (kb) &SMR & Mean& AN50(kb)\\\hline
1 &6.2 &71.5 &254   &6.7 &71.1 &229   &6.2 &72.7 &234  &6.2 &72.7 &234 \\\hline
2 &6.9 &68.6 &241   &8.3 &68.3 &219   &6.9 &69.7 &223  &6.9 &69.7 &223 \\\hline
3 &8.1 &69.7 &218   &8.6 &69.3 &195   &8.0 &70.6 &204  &8.0 &70.6 &204 \\\hline
4 &10.0&63.4 &192   &10.4&63.1 &172   &9.9 &64.6 &177  &9.9 &64.6 &177 \\\hline
5 &8.2 &69.5 &219   &8.8 &69.0 &206   &8.2 &70.3 &210  &8.2 &70.3 &210 \\\hline
6 &7.3 &82.4 &243   &7.9 &81.9 &224   &7.3 &84.0 &236  &7.3 &84.0 &236 \\\hline
7 &7.2 &69.7 &222   &7.6 &69.5 &207   &7.1 &71.0 &212  &7.1 &71.0 &212 \\\hline
8 &7.8 &75.6 &229   &8.3 &75.2 &207   &7.7 &76.8 &220  &7.7 &76.8 &220 \\\hline
9 &7.0 &79.6 &249   &7.5 &79.2 &230   &6.9 &80.9 &235  &6.9 &80.9 &235 \\\hline
10&6.8 &83.9 &238   &7.3 &83.4 &217   &6.7 &84.9 &220  &6.7 &84.9 &220 \\\hline
11&7.1 &77.1 &234   &7.5 &76.8 &225   &7.0 &78.3 &228  &7.0 &78.3 &228 \\\hline
12&6.4 &73.4 &262   &7.3 &73.0 &241   &6.7 &74.1 &249  &6.7 &74.1 &249 \\\hline
13&10.2&69.1 &203   &10.7&68.7 &186   &10.1&70.3 &191  &10.1&70.3 &191 \\\hline
14&6.5 &77.5 &259   &7.0 &77.1 &238   &6.3 &78.4 &246  &6.3 &78.4 &246 \\\hline
15&6.0 &73.7 &251   &6.4 &73.2 &228   &5.9 &74.1 &234  &5.9 &74.1 &234 \\\hline
16&3.8 &96.6 &345   &4.2 &96.2 &317   &3.7 &97.9 &327  &3.7 &97.9 &327 \\\hline
17&3.9 &70.8 &323   &4.5 &70.4 &305   &3.9 &71.5 &310  &3.9 &71.5 &310 \\\hline
18&7.1 &75.3 &228   &7.6 &74.9 &216   &7.0 &76.0 &223  &7.0 &76.0 &223 \\\hline
19&3.1 &90.8 &374   &3.5 &90.4 &345   &3.0 &93.8 &360  &3.0 &93.8 &360 \\\hline
20&4.3 &92.4 &314   &4.8 &92.0 &297   &4.2 &93.7 &304  &4.2 &93.7 &304 \\\hline
21&6.6 &81.1 &252   &7.0 &80.8 &242   &6.4 &82.4 &242  &6.4 &82.4 &242 \\\hline
22&2.7 &123.7&445   &3.2 &123.2&425   &2.6 &123.9&426  &2.6 &123.9&426 \\\hline
\end{tabular}}
\label{table1}
\end{table}

To assess the accuracy of different algorithms, the corresponding reconstruction rates \cite{hapcut, kul} are presented in Fig 2. Moreover, we have considered both short and long SWERs \cite{hapcut, kul}. By a long switch, we mean that the parental origin does not change for at least two SNPs and if two switches occur one after each other, we consider it as a short switch. These two metrics are reported on real fosmid data in Figs 3 and 4.

\begin{figure}[!h]
  \centering
    \includegraphics[width=1\textwidth]{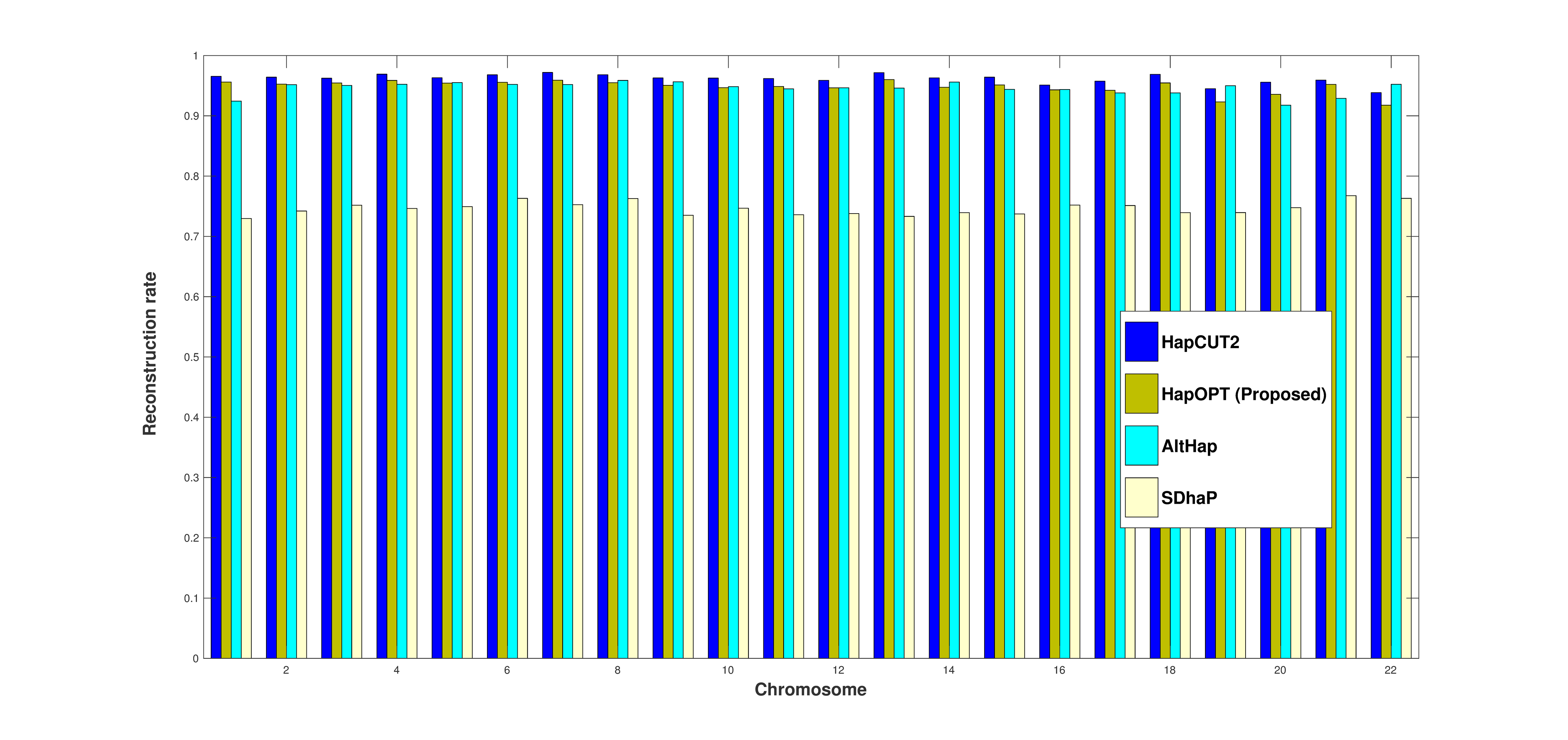}
\caption{{\bf Reconstruction rate of HapOPT, HapCUT2, AltHap, and SDhaP on real fosmid data.}}
\label{fig1}
\end{figure}

\begin{figure}[!h]
  \centering
\caption{{\bf Short SWER of HapOPT, HapCUT2, AltHap, and SDhaP on real fosmid data.}}
\label{fig1}
\end{figure}

\begin{figure}[!h]
  \centering
    \includegraphics[width=1\textwidth]{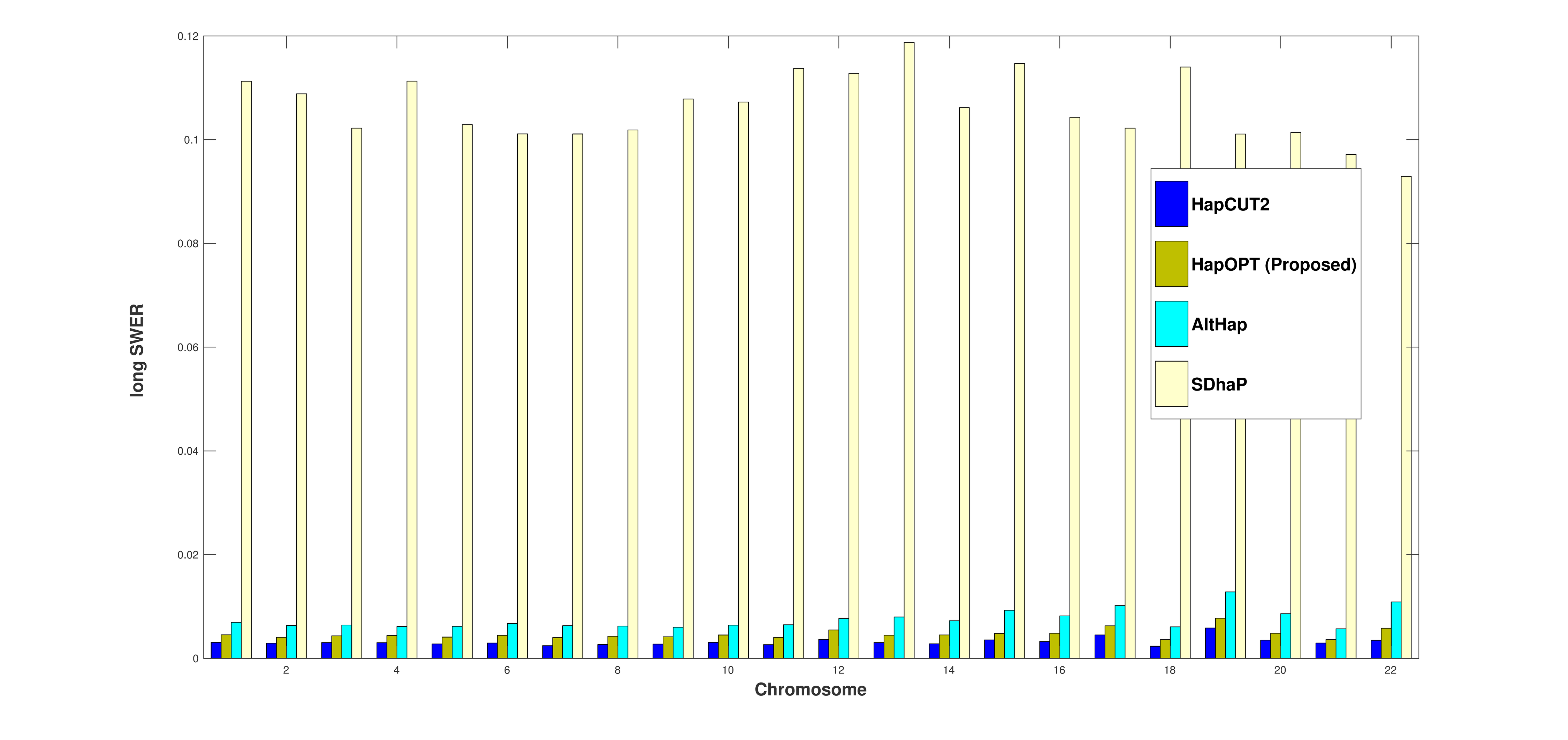}
\caption{{\bf Long SWER of HapOPT, HapCUT2, AltHap, and SDhaP on real fosmid data.}}
\label{fig1}
\end{figure}

\begin{table}[!h]
\centering
\caption{ {\bf Runtime of HapOPT, HapCUT2, AltHap, and SDhaP on real fosmid data.}}
{ \begin{tabular}{|c |c|c|c |c|} \hline
 & SDhaP &AltHap& HapCUT2 & HapOPT (Proposed) \\\hline
Runtime (Minutes) & 5 & 10 &  18  & 355 \\\hline
\end{tabular}}
\label{table1}
\end{table}

  \newpage
From the above results, one can observe that HapOPT outperforms SDhaP and AltHap in terms of the reconstruction rate as well as long and short SWERs with a reasonable runtime as reported in Table 6. Note that although, HapCUT2 achieves the best accuracy, still its SNP missing rate is greater than that of HapOPT. These results on the whole show that  HapOPT is a promising tool for haplotype assembly with the best SNP missing rate and a good accuracy in terms of reconstruction rate and SWER.


\section*{Conclusion}
We have exploited matrix completion methods including SVT, nuclear norm minimization, and OPTSPACE for haplotype estimation. This was led to developing the new HapSVT, HapNuc, and HapOPT algorithms.
Our experimental comparison on simulated data revealed  that HapOPT is more accurate than SDhaP and AltHap in terms of reconstruction rate and switch error rate. Also, the results on real noisy fosmid data showed that the accuracy of HapOPT is better than that of SDhaP and AltHap and also is comparable to that of HapCUT2 in terms of the reconstruction rate and the short and long SWERs. Moreover, it was shown that HapOPT outperforms the recently addressed algorithms, HapCUT2 and SDhaP, in terms of the mean, SNP missing rate, and AN50 of the haplotype block length.
Furthermore, the proposed algorithm is not restricted to the heterozygous assumption, as commonly considered in peer algorithms. On the whole, we can conclude that using the proposed HapOPT, the haplotype is reconstructed more completely and continuously with acceptable accuracy. Also, the proposed optimization problem is capable of estimating haplotypes for different ploidy levels.  Our research direction for future is to work on polyploids.

\subsection*{Availability of data and materials}
The MATLAB program of the proposed algorithms is publicly available at \url{https://github.com/smajidian/HapMC}. The simulated datasets consisting of read matrices and true haplotypes used in this work can be downloaded from  \url{https://github.com/smajidian/HapMC/raw/master/data/Simulated_data.mat.zip}. The fosmid dataset for NA12878 is taken from \cite{kul, dep}. The fragment files can be downloaded from    \url{https://github.com/smajidian/HapMC/raw/master/data/phasing-matrices.zip}  and the grand truth haplotypes are available at   \url{https://github.com/smajidian/HapMC/raw/master/data/validation.zip}.

\end{document}